\documentclass{article}
\usepackage{times}
\usepackage{amsfonts}
\begin{document}
\title{A Quantum--Gravity Perspective\\ on\\ Semiclassical {\it vs.} Strong--Quantum Duality}
\author{Jos\'e M. Isidro\\
Instituto de F\'{\i}sica Corpuscular (CSIC--UVEG)\\
Apartado de Correos 22085, Valencia 46071, Spain\\
{\tt jmisidro@ific.uv.es}}

\maketitle

\begin{abstract}

\noindent

\end{abstract}
\noindent
It has been argued that, underlying M--theoretic dualities, there should exist a symmetry relating the semiclassical and the strong--quantum regimes of a given action integral. On the other hand, a field--theoretic exchange between long and short distances (similar in nature to the T--duality of strings) has been shown to provide a starting point for quantum gravity, in that this exchange enforces the existence of a fundamental length scale on spacetime. In this letter we prove that the above semiclassical {\it vs.}\/ strong--quantum symmetry is equivalent to the exchange of long and short distances. Hence the former symmetry, as much as the latter, also enforces the existence of a length scale. We apply these facts in order to classify all possible duality groups of a given action integral on spacetime, regardless of its specific nature and of its degrees of freedom.

\tableofcontents

\section{Introduction}\label{pajareshijodeputa}

Quantum gravity effects are due to arise at scales of the order of the Planck length $L_P$. Let ${\rm d}s$ denote the Lorentz--invariant interval on a spacetime manifold $\mathbb{M}$. It has been argued that the existence of a fundamental length scale $L_P$ implies modifying the spacetime interval according to the rule
\begin{equation}
{\rm d}s^2\longrightarrow{\rm d}s^2+L_P^2,
\label{ramallo,komprateunchampuantikaspa}
\end{equation}
so $L_P$ effectively becomes the shortest possible distance. In ref. \cite{PADUNO} it has been proved that modifying the spacetime interval according to (\ref{ramallo,komprateunchampuantikaspa}) is equivalent to requiring invariance of a field theory under the following exchange of short and long distances:
\begin{equation}
{\rm d}s\longleftrightarrow\frac{L_P^2}{{\rm d}s}.
\label{paddy}
\end{equation}
This equivalence has been proved in ref. \cite{PADUNO} for the quantum theory of a scalar field; interesting applications of the transformation (\ref{paddy}) to other quantum field theories have been worked out in ref. \cite{PADOS}. Not only field theory: also strings, thanks to T--duality \cite{PORRATI}, are symmetric under the exchange of short and long distances.

Along an apparently unrelated line, section 6 of ref. \cite{VAFA} introduces the concept of {\it duality}\/ as the relativity in the notion of a quantum. In ref. \cite{FEY} we have shown how to complement Feyman's exponential of the action
\begin{equation}
{\rm exp}\left({{\rm i}\,\frac{S}{\hbar}}\right)
\label{ramallocasposo}
\end{equation}
in order to render it manifestly invariant under the exchange
\begin{equation}
\frac{S}{\hbar}\longleftrightarrow\frac{\hbar}{S}.
\label{ramalloleproso}
\end{equation}
Above, the precise nature of the action $S$ is immaterial. It can be, {\it e.g.}, the action for a mechanical system with a finite number of degrees of freedom, or the field theory actions studied in refs. \cite{PADUNO, PADOS}, or any other. The key property of the duality (\ref{ramalloleproso}) is the fact that it maps the semiclassical regime $S\gg\hbar$ into the strong quantum regime $S\ll\hbar$, and viceversa, thus implementing the relativity in the notion of a quantum alluded to above as a duality \cite{VAFA}. 

In this letter we first prove that the dualities (\ref{paddy}) and (\ref{ramalloleproso}) are equivalent: whenever the one holds, so does the other, and viceversa.  We then use this fact in order to provide a geometrical classification of possible duality transformations. We find that the $\mathbb{Z}_2$--transformations (\ref{paddy}) and (\ref{ramalloleproso}) extend to more general groups, among which ${\rm PSL}\,(2,\mathbb{C})$, and subgroups thereof, stand out. Finally we interpret our results in the light of a quantum theory of gravity. Closely related ideas have been put forward in ref. \cite{MARCO}, where Planck's constant $\hbar$ has been interpreted as a dynamically generated quantum scale (which runs according to a certain beta function).

\section{Equivalence of dualities}\label{ramallorijoso}

In this section we establish the equivalence between the dualities (\ref{paddy}) and (\ref{ramalloleproso}). In order to prove this point we digress to introduce the necessary material from ref. \cite{FEY}. 

The generating function for the Bessel functions $J_n(x)$ of integer order $n$ (see, {\it e.g.}, ref. \cite{CH}) is $\exp\left(w(v-v^{-1})/2\right)$:
\begin{equation}
{\rm e}^{\frac{w}{2}(v-\frac{1}{v})}=\sum_{n=-\infty}^{\infty}v^n\,J_n(w), \qquad 0<\vert v\vert<\infty.
\label{pajareschupameelpiton}
\end{equation}
The choice of variables
\begin{equation}
w:=\frac{S}{\hbar}, \qquad v-v^{-1}:= 2{\rm i},
\label{nene}
\end{equation}
gives the expansion
\begin{equation}
{\rm e}^{{\rm i}\frac{S}{\hbar}}=\sum_{n=-\infty}^{\infty}{\rm i}^nJ_n\left(\frac{S}{\hbar}\right).
\label{eodeo}
\end{equation}
The above is an infinite sum of terms, each one of which satisfies the Bessel equation of order $n$,
\begin{equation}
\frac{{\rm d}^2}{{\rm d}w^2}J_n(w)+\frac{1}{w}\frac{{\rm d}}{{\rm d}w}J_n(w)+\left(1-\frac{n^2}{w^2}\right)J_n(w)=0, \qquad n=0,\pm 1,\pm 2, \ldots
\label{ramallomarikonketedenpordetras}
\end{equation}

Some insight into the physical meaning of the expansion (\ref{eodeo}) can be gained from the following observation. Consider the time--independent Schr\"odinger wave equation for a free nonrelativistic particle of mass $m$ on an auxiliary copy of the plane $\mathbb{R}^2$,
\begin{equation}
-\frac{\hbar^2}{2m}\nabla^2\psi=E\psi,
\label{mac}
\end{equation}
which we solve by separation of variables in polar coordinates $r,\varphi$. Then the Laplacian operator $\nabla^2$ on this auxiliary $\mathbb{R}^2$ has the radial piece 
\begin{equation}
\frac{{\rm d}^2}{{\rm d}\rho^2}+\frac{1}{\rho}\frac{{\rm d}}{{\rm d}\rho}+\left(1-\frac{l^2}{\rho^2}\right),\qquad l=0,\pm 1,\pm 2,\ldots,
\label{barbonmekagoentuputabarbacasposa}
\end{equation}
where the radial variable $r$ (with dimensions of length) is related to the dimensionless $\rho$ through
\begin{equation}
\rho:=\lambda^{-1} r, \qquad \lambda:=\frac{\hbar}{\sqrt{2mE}}.
\label{lex}
\end{equation}
The solutions to the radial piece obtained by equating (\ref{barbonmekagoentuputabarbacasposa}) to zero are Bessel functions $J_l(\rho)$ of integer order $l\in\mathbb{Z}$. The eigenfunctions $Y_l(\varphi)$ of the angular momentum operator corresponding to (\ref{mac}) and (\ref{barbonmekagoentuputabarbacasposa}) are
\begin{equation}
Y_l(\varphi)={\rm e}^{{\rm i}l\varphi}, \qquad l=0,\pm 1,\pm 2, \ldots
\label{ramallo,apestas-demierda}
\end{equation}
The general solution to the Schr\"odinger equation (\ref{mac}) is a linear combination 
\begin{equation}
\sum_{l=-\infty}^{\infty}c_l\,J_l(\rho)\,Y_l(\varphi), 
\label{mini}
\end{equation}
the $c_l\in\mathbb{C}$ being certain coefficients.

Now, it is well known that given a classical action $S$ on spacetime $\mathbb{M}$, a semiclassical wavefunction $\psi$ for the corresponding quantum--mechanical problem is obtained as $\psi={\rm e}^{{\rm i}S/\hbar}$, where $S$ satisfies the Hamilton--Jacobi equation \cite{LANDAU}. For a stationary state we can separate out the time dependence ${\rm e}^{-{\rm i}Et/\hbar}$. Consider the auxiliary plane $\mathbb{R}^2$ spanned by the dimensionless  polar coordinates $\rho, \varphi$. The previous argument shows that the $n$--th term in the expansion (\ref{eodeo}) is a partial--wave contribution to the Feynman wave ${\rm e}^{{\rm i}{S}/{\hbar}}$, where the dimensionless radial coordinate $\rho$ on the auxiliary $\mathbb{R}^2$ is identified with the action $S$ as measured in units of the quantum $\hbar$, {\it i.e.},  
\begin{equation}
\rho=\frac{\vert S\vert}{\hbar}=\vert w\vert.
\label{ramallopiojoso}
\end{equation}
To complete the picture, we identify the angular variable $\varphi$ on the auxiliary $\mathbb{R}^2$ as
\begin{equation}
\varphi=2\pi{\rm Re}\left({\rm e}^{-{\rm i}Et/\hbar}\right),
\label{ramallokomprateunchampuantikaspa}
\end{equation}
while the angular momentum $l\in\mathbb{Z}$ in eqn. (\ref{ramallomarikonketedenpordetras}) is identified with the index $n\in\mathbb{Z}$. To summarise, Feynman's time--dependent exponential of the action becomes 
\begin{equation}
{\rm e}^{\frac{{\rm i}}{\hbar}(S-Et)}=\sum_{l=-\infty}^{\infty}{\rm i}^lJ_l\left(\frac{S}{\hbar}\right)\,\exp\left[{2\pi{\rm i}\,l\,{\rm Re}\,\left({\rm e}^{-{\rm i}Et/\hbar}\right)}\right].
\label{ramallo,piojos&grenyas&liendres}
\end{equation}
This is an infinite sum over all possible angular momenta $l\in\mathbb{Z}$ of the auxiliary particle on the auxiliary $\mathbb{R}^2$. As stressed in ref. \cite{FEY}, the auxiliary particle of mass $m$ is not to be confused with the physical degrees of freedom of the action $S$ under consideration in eqn. (\ref{ramallocasposo}). Nor is the auxiliary plane $\mathbb{R}^2$ to be confused with the spacetime $\mathbb{M}$ where $S$ is defined. However, the introduction of this auxiliary particle on $\mathbb{R}^2$ turns out to be a useful device for our purposes.

Initially one may assume that the polar coordinates $\rho,\varphi$ of eqns. (\ref{lex})--(\ref{ramallokomprateunchampuantikaspa}) cover all of the auxiliary $\mathbb{R}^2$ and nothing else. However, there is no reason for $\rho,\varphi$ to be global coordinates. More generally, $\rho, \varphi$ could be {\it local}\/ coordinates on a certain auxiliary surface $\mathbb{S}$ other than $\mathbb{R}^2$. For example, imagine that $\mathbb{S}$ is the Riemann sphere $\mathbb{CP}^1$, and let us consider the local holomorphic coordinate on $\mathbb{CP}^1$ given by
\begin{equation}
z:=\rho\,{\rm e}^{{\rm i}\varphi}.
\label{barbonchupameelbolon}
\end{equation}
Now the point at infinity on $\mathbb{CP}^1$ is not covered by the coordinate (\ref{barbonchupameelbolon}). However we may reach this point by introducing the new holomorphic coordinate $\tilde z$ on $\mathbb{CP}^1$
\begin{equation}
\tilde z:= -\frac{1}{z}=\tilde \rho\,{\rm e}^{{\rm i}\tilde\varphi},
\label{ramallohijoputa}
\end{equation}
where
\begin{equation}
\tilde \rho=\frac{\hbar}{S}, \qquad \tilde\varphi=-(\varphi+\pi).
\label{ramallokambiateloskalzones,almenosunavezporsemana}
\end{equation}
This leads one to the Feynman--like exponential 
\begin{equation}
{\rm exp}\left({\rm i}\,\frac{\hbar}{S}\right)
\label{ramallomariconazo}
\end{equation}
as a candidate for describing the strong quantum regime of a theory whose auxiliary surface $\mathbb{S}$ is $\mathbb{CP}^1$. Then the new choice of variables in eqn. (\ref{pajareschupameelpiton})
\begin{equation}
w:= \frac{\hbar}{S}, \qquad v-v^{-1}:= 2{\rm i}
\label{ramalloduchate,kehuelesamierda}
\end{equation}
leads to the expansion
\begin{equation}
{\rm e}^{{\rm i}\frac{\hbar}{S}}=\sum_{n=-\infty}^{\infty}{\rm i}^nJ_n\left(\frac{\hbar}{S}\right).
\label{barbonketefollen}
\end{equation}
The semiclassical regime of (\ref{barbonketefollen}) is mapped into the strong quantum regime of (\ref{eodeo}), and viceversa. 

{}For a small trajectory of order $\Delta q$, the time--independent piece of the action $S=\int p{\rm d}q$ can be approximated by $p\Delta q$. Under the duality (\ref{paddy}), where $\Delta q\rightarrow L_P^2/\Delta q$, the action transforms as 
\begin{equation}
S=p\Delta q\longrightarrow \frac{pL_P^2}{\Delta q}=\frac{p^2L_P^2}{S}.
\label{padworth}
\end{equation}
Given that $\vert z\vert=\vert S\vert/\hbar$, eqn. (\ref{padworth}) is the dimensionful equivalent of the transformation (\ref{ramallohijoputa}) on $\mathbb{CP}^1$.

To summarise, the statement that the duality (\ref{paddy}) holds is equivalent to the statement that one can transform the coordinate $z$ on the auxiliary surface $\mathbb{S}$ as per eqn. (\ref{ramallohijoputa}). In turn, this latter statement is equivalent to the existence of the duality (\ref{ramalloleproso}) between the semiclassical and the strong--quantum regimes.

\section{A classification of duality transformations}\label{barbonketelametanporkulo}

We have in the foregoing section analysed the cases when the auxiliary surface $\mathbb{S}$ is the plane $\mathbb{R}^2$ and the Riemann sphere $\mathbb{CP}^1$. Let us now be more general and consider an arbitrary auxiliary surface $\mathbb{S}$. Diffeomorphisms of $\mathbb{S}$ that are globally defined are called {\it automorphisms}\/ of $\mathbb{S}$. The set of all automorphisms of $\mathbb{S}$ defines a group, denoted ${\rm Aut}\,(\mathbb{S})$. Elements of ${\rm Aut}\,(\mathbb{S})$ are duality transformations of the physical theory described by the action $S$ whose auxiliary surface is $\mathbb{S}$.

Given a certain duality group $G$ of a given action $S$ on a spacetime $\mathbb{M}$, we will look for an auxiliary surface $\mathbb{S}$ such that ${\rm Aut}\,(\mathbb{S})\subset G$, if possible such that ${\rm Aut}\,(\mathbb{S})=G$.  In this section we perform a partial classification of duality groups. Under {\it partial}\/ we understand that, in general, one will not have ${\rm Aut}\,(\mathbb{S})\subset G$. Rather, in the general case, $G$ will be an extension of ${\rm Aut}\,(\mathbb{S})$.

The geometry of $\mathbb{S}$ will be dictated by the kind of duality transformations that one wishes to implement for the physical action $S$ on spacetime $\mathbb{M}$ \cite{VAFA}. When $\mathbb{S}$ is a complex manifold we will further require that the above--mentioned automorphisms be complex--analytic with respect to the complex structure on $\mathbb{S}$. Riemann surfaces immediately come to mind as possible candidates for the auxiliary surface $\mathbb{S}$; for the rest of this section, a good reference on Riemann surfaces is \cite{FARKAS}. Then a local holomorphic coordinate on $\mathbb{S}$ will be given by eqn. (\ref{barbonchupameelbolon}), with (\ref{ramallopiojoso}) and (\ref{ramallokomprateunchampuantikaspa}) holding true.

{}First and foremost, theories exhibiting no dualities will have $\mathbb{S}=\mathbb{R}^2$. Such is the case of standard quantum mechanics as presented, {\it e.g.}, in ref. \cite{LANDAU}.  
In this case no coordinate transformation is allowed to map the semiclassical regime into the strong--quantum regime, or viceversa. This implies that $\vert z\vert=\vert S\vert/\hbar$ must remain constant. Therefore standard quantum mechanics is described by the {\it real}\/ auxiliary surface $\mathbb{R}^2$ {\it with no complex structure on it}, and its group of allowed automorphisms is the group of isometries of the Euclidean metric ${\rm d}x^2+{\rm d}y^2$ on $\mathbb{R}^2$.

\subsection{The noncompact case}\label{arfonzodemierda}

Consider first the complex plane $\mathbb{C}$, which equals $\mathbb{R}^2$\/ {\it endowed with a complex structure}. The group ${\rm Aut}\,(\mathbb{C})$ is the group of affine transformations  
\begin{equation}
z\rightarrow\tilde z:= az+b, \qquad a,b\in\mathbb{C}, \qquad a\neq 0.
\label{affa}
\end{equation}
It is generated by rotations, translations and dilations. While translations and rotations have no effect on the value of $\vert z\vert$, dilations certainly do.  Thus the complex plane $\mathbb{C}$ corresponds to theories allowing for dualities of the action $S$ on spacetime $\mathbb{M}$.

{}For the upper half--plane $\mathbb{H}:=\left\{z\in\mathbb{C}: {\rm Im}\,z>0\right\}$, the group of automorphisms is ${\rm PSL}\,(2,\mathbb{R})$. Its elements are the transformations
\begin{equation}
z\longrightarrow\tilde z:=\frac{az+b}{cz+d}, \qquad a,b,c,d\in\mathbb{R}, \qquad ad-bc=1.
\label{ramolla,chupamelapolla}
\end{equation}
Again this group allows for changes in the value of $\vert z\vert$. $\mathbb{C}$ and $\mathbb{H}$ by no means exhaust all possible noncompact Riemann surfaces, but let us move on to the compact case.

\subsection{The compact case}\label{arfonzodekaka}

All smooth, compact, connected, closed 2--dimensional manifolds can be classified topologically: any such manifold is homeomorphic either to a sphere with $g$ handles attached to it, or to a sphere with $b$ M\"obius bands attached to it. Now the M\"obius band is nonorientable. Nonorientability implies that one cannot tell between positive and negative values of the angular momentum $l\in\mathbb{Z}$ in eqns. (\ref{ramallo,apestas-demierda}), (\ref{ramallo,piojos&grenyas&liendres}). To avoid this possibility we will concentrate on the case of orientable manifolds $\mathbb{S}$. Then we are left with a sphere with $g$ handles, which is a {\it compact}\/ Riemann surface $\Sigma$ in genus $g$.

\subsubsection{$g=0$}\label{arfonzoderkulo}

In $g=0$ we have $\Sigma=\mathbb{CP}^1$. The latter is invariant under the ${\rm PSL}\,(2,\mathbb{C})$--action
\begin{equation}
z\longrightarrow\tilde z:=\frac{az+b}{cz+d},\qquad a,b,c,d\in\mathbb{C} \qquad ad-bc=1.
\label{ramayo,ketepartaunrayo}
\end{equation}
Therefore  ${\rm Aut}\,\left(\mathbb{CP}^1\right)={\rm PSL}\,(2,\mathbb{C})$. This group is generated by rotations, translations, dilations and the inversion.

\subsubsection{$g=1$}\label{barbondelkulo}

In $g=1$ we have that $\Sigma=T^2$, a complex torus with modular parameter $\tau$, where ${\rm Im}\,\tau>0$. One proves that ${\rm Aut}\,(T^2)$ contains $T^2=\mathbb{C}/H$ as a commutative subgroup. Here $H$ is the group generated by $z\mapsto z+\tau$ and $z\mapsto z+1$, with $z$ a complex coordinate on $\mathbb{C}$. Moreover, for most $\tau$ (in particular, for transcendental $\tau$), the group ${\rm Aut}\,(T^2)$ is a $\mathbb{Z}_2$--extension of $T^2$.

\subsubsection{$g\geq 2$}\label{alvarezgaumekeefollen}

If $g\geq 2$, then ${\rm Aut}\,(\Sigma)$ is a finite group.  The group ${\rm Aut}\,(\Sigma)$ can be faithfully represented on the 1st homology group $H_1(\Sigma)$. Specifically, let ${\rm Sp}\,(2g, \mathbb{Z})$ denote the group of $2g\times 2g$ unimodular matrices that respect the  symplectic pairing between the canonical $\alpha$ and $\beta$ cycles in $H_1(\Sigma)$. Then there is a natural homomorphism
\begin{equation}
h\colon{\rm Aut}\,(\Sigma)\longrightarrow{\rm Sp}\,(2g,\mathbb{Z})
\label{keosdenatodosporkulo}
\end{equation}
which, for $g\geq 2$, is injective. When $\Sigma$ is hyperelliptic, ${\rm Aut}\,(\Sigma)$ can be embedded into ${\rm PSL}\,(2,\mathbb{C})$.

\subsection{Summary}\label{lodicho,alvarezgaume:ketefollenporkulo}

Our analysis can be summarised as follows. When the auxiliary surface $\mathbb{S}$ is one of the following Riemann surfaces $\Sigma$: $\mathbb{C}$, $\mathbb{H}$, $\mathbb{CP}^1$, $T^2$, then its corresponding group of automorphisms is easily identified. In $g\geq 2$ this group is always finite; if, moreover, $\Sigma$ is hyperelliptic, then this group can be embedded into ${\rm PSL}\,(2,\mathbb{C})$. In most cases one has that ${\rm Aut}\,(\Sigma)$ is a subgroup (possibly finite) of ${\rm PSL}\,(2,\mathbb{C})$, and that the physical duality group $G$ is an extension (possibly trivial) of this subgroup of ${\rm PSL}\,(2,\mathbb{C})$. As the notation indicates, ${\rm PSL}\,(2,\mathbb{C})$ is a group of {\it projective}\/ transformations. The projective nature of the corresponding dualities is borne out by eqns. (\ref{ramallopiojoso}), (\ref{ramallokomprateunchampuantikaspa}), (\ref{barbonchupameelbolon}). We conclude that the definition of semiclassical {\it vs.}\/ strong--quantum duality can be taken well beyond the original $\mathbb{Z}_2$--transformation of eqn. (\ref{ramalloleproso}). The latter arises as the inversion within the nonabelian group ${\rm PSL}\,(2,\mathbb{C})$.

\subsection{Examples}

An example where the duality (\ref{ramalloleproso}) arises is the following. Consider the U(1) Born--Infeld Lagrangian
\begin{equation}
{\cal L}_{\rm BI}=\det\left(\eta_{\mu\nu}+bF_{\mu\nu}\right)^{1/2}.
\label{bbii}
\end{equation}
Above, $\eta_{\mu\nu}$ is the Mikowski metric on spacetime and $F_{\mu\nu}$ is the field strength of a U(1)--valued gauge field $A_{\mu}$, while $b$ is a constant. For example, when one couples Born--Infeld electrodynamics to a point particle of mass $m$ and electric charge $e$, we have that $b=e/(ma_{\rm max})$, where $a_{\rm max}$ is the maximal acceleration possible \cite{SCHULLER}. Now, in natural units, the inverse $a_{\rm max}^{-1}$ of a {\it maximal}\/ acceleration is a {\it minimal}\/ length, that we can identify (possibly up to numerical factors) with the {\it Planck length}\/ $L_P$ on spacetime. Hence the existence of a maximal acceleration is equivalent to the existence of a minimal length scale. On the other hand, in section \ref{ramallorijoso} we have proved that the following two statements 
are equivalent:\\
{\it i)} there exists a fundamental length scale $L_P$ on spacetime;\\
{\it ii)} one can perform the exchange (\ref{ramalloleproso}).\\
It follows that the Born--Infeld Lagrangian (\ref{bbii}) exhibits the duality (\ref{ramalloleproso}).

That the Born--Infeld Lagrangian must exhibit the duality (\ref{ramalloleproso}) is easy to understand from an alternative standpoint. The (bosonic piece of the) Lagrangian for branes contains the Born--Infeld term. Moreover, branes are solitonic solutions to the supergravity equations of motion \cite{VAFA}. The latter contain Einstein's equations for the gravitational field. Upon quantisation we expect a fundamental length scale to arise. The duality (\ref{ramalloleproso}) then follows by the previous arguments.

\section{Discussion}\label{gaumekabron}

In this letter we have unveiled a quantum--gravity perspective on the duality between the semiclassical and the strong--quantum regimes corresponding to an action integral $S$ on a spacetime manifold $\mathbb{M}$. That this latter duality should exist was suggested, from a string--theory viewpoint, in ref. \cite{VAFA}. Besides string theory, there are other approaches to a quantum theory of gravity. Together with string theory they all share one common feature, namely, the existence a minimal length scale on spacetime, the Planck length $L_P$.

The duality (\ref{paddy}) arises naturally in the geometrical setup of quantum gravity. On the contrary its close cousin (\ref{ramalloleproso}), though equivalent, may on first sight appear puzzling. After all, Planck's quantum of action $\hbar$ is a fundamental constant in units of which all physical observables with a discrete spectrum are quantised, while the duality between short and long distances is of a more geometrical nature. It may thus cause some concern to even consider physical processes in which the measurable action $S$ becomes much smaller than the quantum of action $\hbar$. However, a moment's reflection shows that similar objections might be raised to the transformation (\ref{paddy}), given that $L_P$ is also a fundamental constant for every given spacetime dimension. Therefore, if it makes sense to consider the {\it geometrical}\/ duality (\ref{paddy}), it makes no less sense to consider its close cousin, the {\it physical}\/ duality (\ref{ramalloleproso}). Then the same reasoning that led one to require invariance under (\ref{paddy}) will also apply to require invariance under (\ref{ramalloleproso}). 

The existence of the Planck scale has been shown to be equivalent to the requirement that the duality (\ref{paddy}) hold true \cite{PADUNO, PADOS}. In this letter we have established the equivalence between the duality (\ref{paddy}) and that proposed in ref. \cite{FEY}, which can be summarised in eqn. (\ref{ramalloleproso}). 

We have in ref. \cite{ME} advocated an approach to quantum gravity based on an attempt to render the notion of a quantum relative to the observer. In this dual approach, a prominent role is played by the auxiliary surface $\mathbb{S}$. The latter is introduced as a real 2--dimensional manifold spanned by (certain functions of) the dynamical variables $S$ and $Et$. As already explained, our viewpoint can be summarised in the statement that rendering the notion of a quantum {\it relative}\/ is dual to quantising the theory of relativity. Here the term {\it relative}\/ refers to the fact that there is no preferred location and/or preferred choice of coordinates on the auxiliary surface $\mathbb{S}$.

That rendering the notion of a quantum {\it relative}\/ provides a dual approach to a quantum theory of relativity is more than just a pun on words. The above line of reasoning
establishes the following chain of equivalences: by ref. \cite{PADUNO}, the existence of a fundamental length scale $L_P$ is equivalent to invariance under
the duality (\ref{paddy}). In turn, the latter is equivalent to the duality (\ref{ramalloleproso}). Hence the semiclassical {\it vs.} strong--quantum duality (\ref{ramalloleproso}) is equivalent to the existence of the Planck length. The existence of a fundamental length scale is a hallmark of any quantum theory of gravity.

Along the way to the above conclusions we have learnt how to extend the original $\mathbb{Z}_2$--transformation (\ref{ramalloleproso}) to more general transformation groups.
The group $G$  of physical dualities of the theory defined by the action $S$ on spacetime $\mathbb{M}$ will be an extension of the group of automorphisms ${\rm Aut}\,(\mathbb{S})$. When this extension is trivial one has $G={\rm Aut}\,(\mathbb{S})$, and our procedure immediately allows one to identify $G$. At worst, one will have to look for all possible extensions of ${\rm Aut}\,(\mathbb{S})$ in order to identify the group of dualities of the given action $S$. 

A substantial simplification is achieved when $\mathbb{S}$ is a Riemann surface $\Sigma$: one can then relatively easily identify the corresponding group of automorphisms ${\rm Aut}\,(\Sigma)$ \cite{FARKAS}. In most of the cases analysed in section \ref{barbonketelametanporkulo} one finds that ${\rm Aut}\,(\Sigma)$ equals (a subgroup  of) ${\rm PSL}\,(2,\mathbb{C})$. The group ${\rm PSL}\,(2,\mathbb{C})$ (and subgroups thereof) is ubiquitous in the presence of conformal symmetry. Thus, {\it e.g.}, the relevance of Riemann surfaces to the quantum theory of 2--dimensional gravity was shown long ago; conformal techniques have been successfully applied more recently in ref. \cite{MATONE} to a variety of related problems. 

Our classification of the possible duality groups makes no reference to the degrees of freedom present in the action $S$. This is so because our treatment makes no assumptions concerning the precise nature of $S$. In particular, our conclusions apply equally well to a finite number of degrees of freedom (a mechanical setup) and to an infinite number of them (a field--theory setup). However, one would expect these degrees of freedom to play a prominent role in the determination of the corresponding duality transformations. The influence that $S$ and the variables it depends on may have on the possible dualities is encoded in the auxiliary surface $\mathbb{S}$. For this reason one would like to have a criterion according to which, given a certain action $S$ on a certain spacetime $\mathbb{M}$, one could determine the corresponding auxiliary surface $\mathbb{S}$. 

The previous arguments also lead to the following conclusion. We have been able to trade the information contained in the length scale $L_P$ on spacetime $\mathbb{M}$ for the information contained in the auxiliary surface $\mathbb{S}$ and its group of automorphisms. That is, given the one piece of information we can recover the other piece, and viceversa. However one is naturally inclined to believe that spacetime $\mathbb{M}$ is more fundamental than the surface $\mathbb{S}$. After all, the latter has been termed {\it auxiliary}. Without elevating the surface $\mathbb{S}$ to the category of a fundamental concept, one can perhaps cut spacetime $\mathbb{M}$ down to measure, if it ceases to be as fundamental a concept as it is usually claimed to be. In fact M--theory has been argued to be a {\it pregeometrical}\/ theory, in that it does not postulate spacetime as an {\it a priori}\/ concept.  Modern theories of quantum gravity also tend to do away with the spacetime continuum as a starting point, only to recover it as a {\it derived}\/ notion, not a primary one. In treating the action $S$ independently of its specific degrees of freedom and in analysing its properties through those of $\mathbb{S}$, regardless of the spacetime $\mathbb{M}$ considered, our conclusions also point in this direction.

{\bf Acknowledgements}

It is a great pleasure to thank Albert--Einstein--Institut (Potsdam, Germany) for hospitality during the early stages of this article. This work has been partially supported by EU network MRTN--CT--2004--005104, by research grant BFM2002--03681 from Ministerio de Ciencia y Tecnolog\'{\i}a, by research grant GV2004--B--226 from Generalitat Valenciana, by EU FEDER funds and by Deutsche Forschungsgemeinschaft.

\end{document}